\documentclass[conference]{IEEEtran}
\IEEEoverridecommandlockouts

\usepackage{cite}
\usepackage{amsmath,amssymb,amsfonts,amsthm}
\usepackage{graphicx}
\usepackage{textcomp}
\usepackage{xcolor}
\def\BibTeX{{\rm B\kern-.05em{\sc i\kern-.025em b}\kern-.08em
    T\kern-.1667em\lower.7ex\hbox{E}\kern-.125emX}}

\usepackage[english]{babel}
\usepackage[T1]{fontenc}
\usepackage[cmintegrals]{newtxmath}
\usepackage{bm}
\usepackage[caption=false,font=footnotesize,farskip=0pt]{subfig}
\usepackage{enumitem}
\usepackage{soul}
\usepackage{booktabs}
\usepackage{multirow}
\usepackage{pgfplots}
\usepackage{hyperref}
\usepackage{balance}

\usepackage[acronym]{glossaries}
\glsdisablehyper
\newacronym{3gpp}{3GPP}{3rd Generation Partnership Project}
\newacronym{5g}{5G}{the 5th generation of mobile networks}
\newacronym{6g}{6G}{sixth generation of mobile networks}
\newacronym{adas}{ADAS}{advanced driver-assitance systems}
\newacronym{ai}{AI}{artificial intelligence}
\newacronym{awgn}{AWGN}{additive white Gaussian noise}
\newacronym{b5g}{B5G}{5G and Beyond}
\newacronym{ber}{BER}{bit error rate}
\newacronym{bler}{BLER}{block error rate}
\newacronym{bs}{BS}{base station}
\newacronym{cdf}{CDF}{cumulative distribution function}
\newacronym{cml}{CML}{commercial microwave link}
\newacronym{crb}{CRB}{Cramér-Rao bound}
\newacronym{crx}{CRx}{cognitive receiver}
\newacronym{csi}{CSI}{channel state information}
\newacronym{ctx}{CTx}{cognitive transmitter}
\newacronym{dl}{DL}{downlink}
\newacronym{dmrab}{DMRAB}{disjoint matching and resource allocation benchmark}
\newacronym{embb}{eMBB}{Enhanced Mobile Broadband}
\newacronym{gan}{GAN}{generative adversarion network}
\newacronym{gap}{GAP}{Generalized Assignment Problem}
\newacronym{geo}{GEO}{Geosynchronous Earth Orbit}
\newacronym{gsl}{GSL}{Ground-to-Satellite Link}
\newacronym{harq}{HARQ}{hybrid automatic repeat request}
\newacronym{ic}{IC}{interference channel}
\newacronym{iot}{IoT}{Internet of Things}
\newacronym{isac}{ISAC}{integrated sensing and communications}
\newacronym{isl}{ISL}{Inter-Satellite Link}
\newacronym{jmra}{JMRA}{joint matching and resource allocation}
\newacronym{jpeg}{JPEG}{joint photographic experts group}
\newacronym{jscc}{JSCC}{joint source channel coding}
\newacronym{kpi}{KPI}{key performance indicator}
\newacronym{ldpc}{LDPC}{low-density parity-check}
\newacronym{leo}{LEO}{Low Earth Orbit}
\newacronym{lu}{LU}{licensed user}
\newacronym{m-pam}{$M$-PAM}{$M$-ary pulse amplitude modulation}
\newacronym{mcs}{MCS}{modulation and coding scheme}
\newacronym{mgap}{MGAP}{Multi-Level Generalized Assignment Problem}
\newacronym{milp}{MILP}{mixed-integer linear problem}
\newacronym{mimo}{MIMO}{Multiple-Input Multiple-Output}
\newacronym{miou}{mIoU}{mean intersection over union}
\newacronym{ml}{ML}{machine learning}
\newacronym{mle}{MLE}{maximum likelihood estimator}
\newacronym{mse}{MSE}{mean-squared error}
\newacronym{nmse}{NMSE}{normalized mean-squared error}
\newacronym{noma}{NOMA}{non-orthogonal multiple access}
\newacronym{nr}{NR}{New Radio}
\newacronym{ntn}{NTN}{Non-Terrestrial Network}
\newacronym{ofdma}{OFDMA}{Orthogonal Frequency-Division Multiple Access}
\newacronym{pdf}{PDF}{probability density function}
\newacronym{pdv}{PDV}{packet delay variation}
\newacronym{pmf}{PMF}{probability mass function}
\newacronym{png}{PNG}{portable network graphics}
\newacronym{ppp}{PPP}{Poisson point process}
\newacronym{psnr}{PSNR}{peak signal-to-noise ratio}
\newacronym{psv}{PSV}{probability of simultaneity violation}
\newacronym{pu}{PU}{primary user}
\newacronym{qos}{QoS}{Quality of Service}
\newacronym{ra}{RA}{resource allocation}
\newacronym{ran}{RAN}{radio access network}
\newacronym{rssi}{RSSI}{received signal strength indicator}
\newacronym{rtt}{RTT}{round-trip time}
\newacronym{rv}{RV}{random variable}
\newacronym{snr}{SNR}{signal-to-noise ratio}
\newacronym{sr}{SR}{scheduling request}
\newacronym{ss}{SS}{synchronization signal}
\newacronym{su}{SU}{secondary user}
\newacronym{tdd}{TDD}{time-division duplexing}
\newacronym{twi}{TWI}{temporal window of integration}
\newacronym{uav}{UAV}{unmanned aerial vehicle}
\newacronym{ue}{UE}{User Equipment}
\newacronym{ul}{UL}{uplink}
\newacronym{vq-vae}{VQ-VAE}{vector-quantized variational autoencoder}
\newacronym{v2v}{V2V}{vehicle-to-vehicle}
\newacronym{semcom}{SemCom}{Semantic Communications}

\newtheorem{definition}{Definition}

\usepackage{stfloats}
\fnbelowfloat

\usepgfplotslibrary{fillbetween}

\pgfplotsset{
    compat=1.18,
    tick label style={font=\scriptsize},
    label style={font=\scriptsize},
    legend style={font=\scriptsize,draw=none,row sep=-2pt,inner sep=0,outer sep=0,fill=none},
    tick style={color=black},
    major tick length=3pt,
    minor tick length=1.5pt,
    label shift=-4pt
}

\definecolor{vqvae_color}{RGB}{228,26,28}
\definecolor{png_color}{RGB}{55,126,184}
\definecolor{jpeg_color}{RGB}{77,175,74}
\definecolor{rgb_color}{RGB}{152,78,163}

\DeclareMathOperator*{\argmin}{arg\,min}

\begin{document}

\bstctlcite{bst_config}

\title{Low-Latency Task-Oriented Image Transmission with Opportunistic Spectrum Access
\thanks{This work was supported by the SNS JU project 6G-GOALS under the EU’s Horizon Europe program under Grant Agreement No 101139232. The work by J. H. Inacio de Souza and P. Popovski was also supported by the Villum Investigator Grant “WATER” from the Velux Foundation, Denmark.}}

\author{\IEEEauthorblockN{João Henrique Inacio de Souza\IEEEauthorrefmark{1}, Mattia Merluzzi\IEEEauthorrefmark{2}, Mateus P. Mota\IEEEauthorrefmark{2}, Beatriz Soret\IEEEauthorrefmark{3}\IEEEauthorrefmark{1}, Petar Popovski\IEEEauthorrefmark{1}}
\IEEEauthorblockA{\IEEEauthorrefmark{1}\textit{Department of Electronic Systems}, \textit{Aalborg University}, Aalborg, Denmark. E-mail: \{
\pdfstartlink
    attr{/Border [0 0 0]}
    user{/Subtype /Link /A << /S /URI /URI (mailto:jhids@es.aau.dk) >>}%
    jhids%
\pdfendlink
,%
\pdfstartlink
    attr{/Border [0 0 0]}
    user{/Subtype /Link /A << /S /URI /URI (mailto:petarp@es.aau.dk) >>}%
    petarp%
\pdfendlink
\}@es.aau.dk}
\IEEEauthorblockA{\IEEEauthorrefmark{2}\textit{CEA-Leti}, \textit{Université Grenoble Alpes}, F-38000 Grenoble, France. E-mail: \{
\pdfstartlink
    attr{/Border [0 0 0]}
    user{/Subtype /Link /A << /S /URI /URI (mailto:mattia.merluzzi@cea.fr) >>}%
    mattia.merluzzi%
\pdfendlink
,%
\pdfstartlink
    attr{/Border [0 0 0]}
    user{/Subtype /Link /A << /S /URI /URI (mailto:mateus.pontesmota@cea.fr) >>}%
    mateus.pontesmota%
\pdfendlink
\}@cea.fr}
\IEEEauthorblockA{\IEEEauthorrefmark{3}\textit{Telecommunications Research Institute}, \textit{Universidad de Málaga}, Málaga, Spain. E-mail: bsa@uma.es}}

\maketitle

\begin{abstract}
Communication systems designed for reliable data reconstruction, rather than task-oriented communication, typically rely on separate source and channel coding and incur high latency under limited spectrum availability and fading channels. To address this, we propose a transmission framework with opportunistic spectrum access, in which the transmitter sends discrete latent representations learned via a vector-quantized variational autoencoder~(VQ-VAE) over idle licensed channels using standard digital modulation. The AI-powered receiver is still able to reconstruct task-related information from the heavily compressed data. We develop a cross-layer latency model that accounts for compression, block errors, retransmissions, and stochastic channel access. Results on latency-accuracy trade-offs show that the proposed scheme achieves at least 79- and 3.3-fold latency reductions with only 5.7\% and 2.4\% drops in classification accuracy compared to benchmarks using conventional source and channel coding. The framework enables low-latency communication and reliable task execution even under limited spectrum availability and challenging channel conditions.
\end{abstract}


%
%
\section{Introduction}

The convergence of wireless communications and \gls{ai} has led to the paradigm of task-oriented communication, where the objective is to efficiently support downstream tasks rather than to reliably reconstruct transmitted data~\cite{Strinati2024}. In this context, performance is measured through metrics that jointly capture task accuracy and communication efficiency, which is particularly relevant for computer vision applications such as image classification, object detection, and scene understanding.

For many emerging applications, communication must support latency-sensitive tasks, where end-to-end delay is the primary performance metric. Achieving low latency requires reducing the communication payload while preserving task-relevant information, which challenges conventional transmission schemes employing separate source and channel coding.

Recent works on \gls{jscc} have shown that learned representations can provide compact and robust data compression tailored to downstream tasks~\cite{Zhang2024,Xie2023,Hu2023}. In particular, \gls{vq-vae}~\cite{VanDenOord2017} enables compression into a discrete latent space, allowing the resulting representations to be transmitted using standard digital modulation schemes and integrated into link adaptation frameworks. Moreover, such representations have been shown to preserve task-relevant features, enabling competitive classification performance even under compression and channel-induced distortions~\cite{Zhang2024}.

Most existing studies assume dedicated or continuously available communication resources. However, in cognitive radio systems, transmission occurs through opportunistic access to licensed channels, where availability depends on primary user activity~\cite{mitola1999cognitiveradio,haykin2005CR}. This introduces an additional source of randomness, as communication opportunities are intermittent. Consequently, the communication latency depends not only on compression and transmission reliability, but also on channel availability, making its characterization more involved.

\begin{figure}[t]
    \centering
    \includegraphics[width=\linewidth]{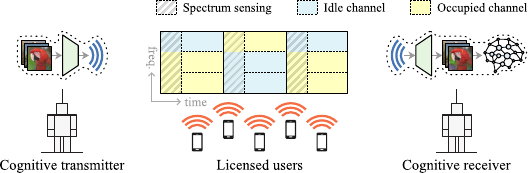}
    \caption{System model for low-latency image transmission with opportunistic spectrum access for \gls{ai}-based remote inference.}
    \label{fig:system-model}
\end{figure}

In this work, we consider a pair of cognitive users that exchange images for remote inference under opportunistic spectrum access. System performance is evaluated in terms of task accuracy and communication latency, defined as the time required to deliver a batch of images for processing. To enable low-latency communication, we adopt a \gls{vq-vae}-based transmission framework that generates discrete latent representations, reducing the number of transmitted bits while remaining compatible with digital communication pipelines.

Our main contribution is a statistical analysis of the communication latency for task-oriented data transmission with opportunistic spectrum access. The analysis captures the impact of compression, block errors, retransmissions, and stochastic channel access, and applies to both the proposed framework and standard benchmarks. It enables the evaluation of latency-accuracy trade-offs in scenarios with limited and unreliable spectrum access. Applying this analysis, we show that the proposed framework achieves communication latency gains of at least $79\times$ and $3.3\times$ with marginal task-accuracy drops of 5.7\% and 2.4\% compared to standard schemes using lossless and lossy source compression, respectively.

The rest of the paper is organized as follows. Section~\ref{sec:system-model} presents the system model. Sections~\ref{sec:vae} and~\ref{sec:comms} describe the proposed framework and the analysis of the communication latency, respectively. Numerical results are discussed in Section~\ref{sec:results}, and Section~\ref{sec:conclusion} concludes the paper. 


\begin{figure*}[b]
    \centering
    \includegraphics[width=\linewidth]{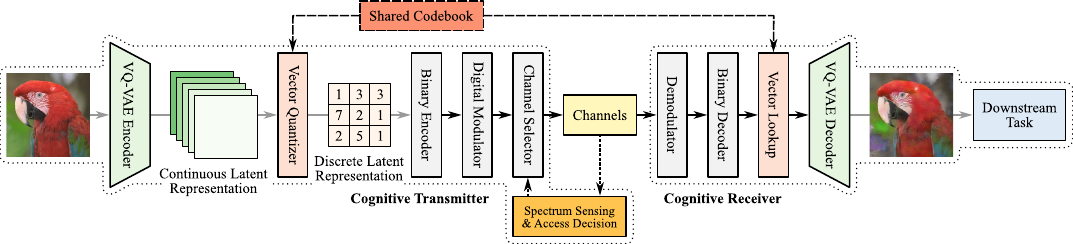}
    \caption{Proposed framework for task-oriented, low-latency image transmission with opportunistic spectrum access, leveraging low-dimensional latent representations learned via a \gls{vq-vae}.}
    \label{fig:proposed-scheme}
\end{figure*}

%
%
\section{System Model}\label{sec:system-model}

We consider the scenario in Fig.~\ref{fig:system-model}, where a \gls{ctx} sends a batch of data samples for remote inference to a \gls{crx}. Both \gls{ctx} and \gls{crx} have a single antenna and use slotted communication over $C$ licensed channels of total bandwidth $B$, with frame duration $T_0$. \Glspl{lu} are willing to share part of their radio resources with cognitive users. The \gls{ctx} performs spectrum sensing to detect \gls{lu} activity on each channel and access idle ones. This is performed at the start of every frame, leaving $T<T_0$ for data transmission. To model spectrum availability accounting for the \gls{lu} activity and sensing performance, we define the channel non-access probability for the \gls{ctx}.

\begin{definition}[Channel non-access probability]\label{def:idle-probability}
    We denote by $p_c$ the probability that the \gls{ctx} does not send data over channel $c$ in a given frame due to the detection of \gls{lu} activity. Regardless of the spectrum sensing method, this probability depends on the method's true and false positive rates, as well as on the distribution of \gls{lu} activity. For simplicity, we assume stationary processes and that \gls{lu} activity is i.i.d. across frames, so that $p_c$ remains constant over time.
\end{definition}

Each data sample is an RGB image $\bm{S} \in \mathbb{R}^{H \times W \times 3}$, where $H$ and $W$ are the image height and width in pixels, and 3 is the number of color components. An uncompressed communication batch is denoted by $\bm{\Sigma} \in \mathbb{R}^{S\times H \times W \times 3}$ with $S$~samples, and it represents the minimum data unit for the \gls{crx} to start inference. The batch contains $b_0$ bits, compressed to $b<b_0$ bits after source compression, giving a compression ratio of $b_0/b$. The \gls{ctx} splits the $b$-bit batch into $n$-bit data blocks, each carrying $k\leq n$ information bits; $k=n$ if uncoded, while $k<n$ when channel coding is used.

Before transmission over physical channels, the data blocks are mapped to symbols of a \gls{m-pam} constellation. We assume block-fading channels with a dominant line-of-sight component and perfect power control. Therefore, the channels are equivalent additive white Gaussian noise channels with \gls{snr} $\gamma$, where \gls{ber} is approximated by~\cite{Proakis2008}
\begin{equation}
    P_b \approx \frac{2(M-1)}{M \log_2 M} Q \left( \sqrt{\frac{6}{M^2-1} \gamma} \right),
\end{equation}
and $Q$ denotes the Gaussian Q-function.

For uncoded transmissions, the \gls{bler} for $n$-bit blocks is approximated as
\begin{equation}
    \epsilon_0 \approx 1 - (1 - P_b)^n,
\end{equation}
assuming independent bit errors. With channel coding, the \gls{bler} $\epsilon$ depends on $k$, $n$, and the coding scheme, and is typically lower than $\epsilon_0$. To trade communication latency for reliability, the \gls{ctx} can enable or disable retransmissions of erroneously received data blocks and the corresponding feedback signaling. When enabled, we assume that the \gls{crx} can perfectly detect block errors and send instantaneous, error-free, per-block requests that result in immediate retransmissions.

In what follows, we introduce the proposed framework for low-latency image transmission based on \gls{vq-vae}.

%
%
\section{Low-Latency Image Transmission via VQ-VAE}\label{sec:vae}

In this section, we introduce the proposed transmission framework as shown in Fig.~\ref{fig:proposed-scheme}. To enable low-latency communication, the framework relies on compressed image representations obtained by a \gls{vq-vae} trained offline. The learned encoder and decoder networks are deployed in the \gls{ctx} and \gls{crx}, respectively, while the learned latent space functions as a codebook shared between the cognitive users.

To transmit data, the \gls{ctx} passes an image through the \gls{vq-vae} encoder and quantizer, obtaining a discrete representation based on latent vectors in the codebook. The indices of the latent vectors, which are scalars in the domain $[1,K]$, are then binary-encoded as sequences of $\log_2 K$ bits, and the resulting bitstream is mapped to \gls{m-pam} symbols. Finally, the symbols are sent over the idle channels detected during spectrum sensing. Subsequently, to receive the data, the \gls{crx} applies the inverse operations to the received symbols, recovering the discrete representation and feeding it to the \gls{vq-vae} decoder to obtain the reconstructed image. This operation is performed for all samples in the communication batch, and the reconstructed images are forwarded to the downstream task.

\subsection{Vector-Quantized Variational Autoencoder}

Now, we present the \gls{vq-vae} model used to obtain the compressed image representations. Importantly, during offline training, the model is trained in the absence of the wireless communication pipeline. Evaluations of communication latency and task accuracy will show later that this approach leads to stable learning and competitive performance during system operation.

Given an input image $\bm{S}$, the encoder network maps it into a continuous latent representation as:
\begin{equation}
    \bm{Z}_e = f_{\theta}(\bm{S}), \quad \bm{Z}_e \in \mathbb{R}^{H' \times W' \times \Delta},
\end{equation}
where $f_{\theta}$ denotes the encoder parameterized by $\theta$, $H'\times W'$ is the spatial dimension of the latent representation, and $\Delta$ is the dimensionality of the latent space. Each spatial location $ \left(i,j \right)$ of the tensor $\bm{Z}_e$ is then quantized using a learned discrete latent space $\mathcal{E} = \{ \bm{e}_l \in \mathbb{R}^\Delta \}^{K}_{l=1}$, resulting in a discrete latent representation $\bm{Z}_q \in \mathbb{R}^{H' \times W' \times \Delta}$:
\begin{equation}
    \bm{Z}_q^{(i,j)} = \argmin_{\bm{e}_l \in \mathcal{E}} \| \bm{Z}_e^{(i,j)} - \bm{e}_l \|_2.
\end{equation}
Naturally, the quantized feature map $\bm{Z}_q$ can be equivalently represented by a matrix of indices:
\begin{equation}
    \bm{I}^{(i,j)} = \argmin_{l \in \{1,\dots,K\}} \| \bm{Z}_e^{(i,j)} - \bm{e}_l \|_2,
\end{equation}
which, during system operation, can be transmitted more efficiently than the full latent vectors. Hence, since each sample is compressed to $H'W'$ indices, each encoded using $\log_2 K$ bits, the resulting compressed communication batch has $b = H'W'\log_2 K$ bits.

To reconstruct the compressed image, the indices are mapped back to their corresponding latent vectors using $\mathcal{E}$, yielding the latent discrete representation:
\begin{equation}
    \hat{\bm{Z}}_q^{(i,j)} = \bm{e}_{\hat{\bm{I}}^{(i,j)}},
\end{equation}
where $\hat{\bm{Z}}_q$ and $\hat{\bm{I}}$ are used to emphasize that during system operation, reconstruction may rely on partially erroneous received data. Finally, the decoder network $g_{\phi}$, parameterized by $\phi$, produces the reconstructed image as $\hat{\bm{S}} = g_{\phi} ( \hat{\bm{Z}}_e )$.

The loss function of \gls{vq-vae} consists of three terms and is defined as follows~\cite{VanDenOord2017}, where $\text{sg}\{\cdot\}$ denotes the stop-gradient operator:
\begin{align}
    \mathcal{L} = - \log\Pr(\bm{S}|\bm{Z}_q) & + \sum_{(i,j)} \|\text{sg}\{\bm{Z}_e^{(i,j)}\} - \bm{e}_{\bm{I}^{(i,j)}}\|_2^2~+\\
    & + \beta \sum_{(i,j)} \|\bm{Z}_e^{(i,j)} - \text{sg}\{\bm{e}_{\bm{I}^{(i,j)}}\}\|_2^2, \nonumber
\end{align}
where $(i,j) \in H'\times W'$. The first term is the variational lower bound with constant terms omitted, which reduces to the mean-squared error between $\bm{S}$ and $\hat{\bm{S}}$~\cite{VanDenOord2017}. The second term is the vector quantization loss, which moves the latent space vectors toward the encoder outputs. Meanwhile, the third term is the commitment loss weighted by $\beta$, which encourages the encoder to generate outputs that are close to the latent space vectors. By minimizing this loss during offline, unsupervised training, we obtain the discrete latent space and the encoder and decoder networks used for low-latency image transmission.

%
%
\section{Communication Latency Analysis}\label{sec:comms}

In this section, we statistically characterize the communication latency of the investigated system for opportunistic wireless transmission over licensed channels. The characterization considers the general setup and is applicable to analyze the latency of the proposed framework, as well as other transmission schemes that employ, \emph{e.g.}, separate source and channel coding, with retransmissions enabled or disabled.

To model the communication latency, we first derive the number of block transmissions required to send a batch of samples, including retransmissions due to block errors. We then derive the number of channel accesses required to deliver these blocks, accounting for the spectrum availability constrained by \gls{lu} activity. Hence, given the compressed batch of $b$ bits and the block structure with $k$ information bits, the resulting number of blocks is $D_t = \lceil b/k \rceil$.
%

During reception, each data block is subject to a \gls{bler} $\epsilon$. In this context, when retransmissions are enabled, erroneously received blocks are detected and retransmitted, thereby increasing the communication latency. In this case, the number of retransmissions follows a negative binomial distribution of the form $\text{NB}(D_t,1-\epsilon)$. Consequently, depending on whether retransmissions are enabled or disabled, the number of retransmitted blocks is modeled as
\begin{equation}
    \begin{cases}
        D_r \sim \text{NB}(D_t,1-\epsilon), & \hspace{-2mm}\text{if retransmissions are enabled}\\
        D_r = 0, & \hspace{-2mm}\text{if retransmissions are disabled}
    \end{cases}.
\end{equation}
This gives a total of $D_t+D_r$ block transmissions per batch. Accordingly, as each block comprises $n$ bits, the \gls{ctx} must send $(D_t+D_r)n$ bits over the idle channels in order to deliver an entire batch of samples.

Since communication is carried out over time-constrained frames, a batch transmission might span across multiple frames. Hence, given that each channel access supports the transmission of $BTC^{-1} \log_2 M$ bits, the number of accesses required to complete a batch transmission is given by
\begin{equation}
    F_a = \left\lceil \frac{(D_t+D_r)n}{BTC^{-1} \log_2 M} \right\rceil.
\end{equation}
The number of idle channels detected per frame depends on the \gls{lu} activity and the spectrum sensing performance, meaning that the \gls{ctx} might achieve fewer than $C$ channel access decisions per frame interval. Given that, let $F_n$ denote the number of channel non-access decisions made by the \gls{ctx} before making $F_a$ access decisions. $F_n$ depends on $F_a$, $C$, and the channel non-access probabilities $\{p_c\}_{c=1}^C$ as in Definition~\ref{def:idle-probability}. In particular, assuming identical non-access probabilities across channels, \emph{i.e.}, $p_1=\dots=p_C=p$, $F_n$ can be defined such that $F_n \sim \text{NB}(F_a,1-p)$. 
%
%
Then, given that the system has $C$ channels and each frame has a duration $T_0$, the latency to transmit an entire batch is given by
\begin{equation}\label{eq:latency}
    L = \left\lceil \frac{F_a+F_n}{C} \right\rceil T_0.
\end{equation}

As noted earlier, this latency analysis applies either to a system with separate source and channel coding or to the proposed transmission framework. In the case of the proposed framework, we set $n = k = \log_2 K$ and retransmissions are disabled, resulting in $D_r=0$. In the next section, we use the model in \eqref{eq:latency} to evaluate the latency of the proposed framework and the benchmarks.

%
%
\section{Numerical Results}\label{sec:results}

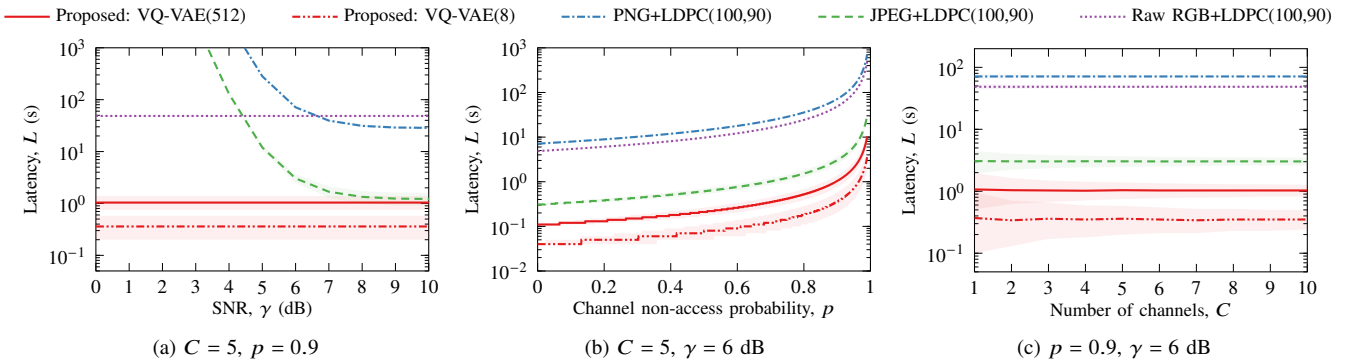
\begin{figure*}[b]
    \centering
    \subfloat{%
    \begin{tikzpicture}
    \begin{axis}[
        width=0cm,
        height=0cm,
        axis line style={draw=none},
        tick style={draw=none},
        at={(0,0)},
        scale only axis,
        xmin=0, xmax=1, xtick={},
        ymin=0, ymax=1, ytick={},
        axis background/.style={fill=white},
        legend style={at={(0, 0)}, anchor=center, /tikz/every even column/.append style={column sep=1.5em}, legend columns=-1},
    ]
    \addlegendimage{draw=vqvae_color,thick}
    \addlegendentry{Proposed: VQ-VAE(512)}
    \addlegendimage{draw=vqvae_color,thick,densely dashdotdotted}
    \addlegendentry{Proposed: VQ-VAE(8)}
    \addlegendimage{draw=png_color,thick,densely dashdotted}
    \addlegendentry{PNG+LDPC(100,90)}
    \addlegendimage{draw=jpeg_color,thick,densely dashed}
    \addlegendentry{JPEG+LDPC(100,90)}
    \addlegendimage{draw=rgb_color,thick,densely dotted}
    \addlegendentry{Raw RGB+LDPC(100,90)}
    \end{axis}
    \end{tikzpicture}
    }
    \\
    \setcounter{subfigure}{0}
    \subfloat[$C=5$, $p=0.9$]{%
    \begin{tikzpicture}
    \begin{semilogyaxis}[
        width=0.33\linewidth,
        height=0.25\linewidth,
        xlabel={SNR, $\gamma$ (dB)},
        ylabel={Latency, $L$ (s)},
        xmin={0}, xmax={10}, xtick distance={1},
        ymin={5e-2}, ymax={1e3}, try min ticks log={5},
    ]
        \addplot[png_color,thick,densely dashdotted] table[col sep=comma,x index=0,y index=1] {data/latency_snr_png.dat};
        \addplot[draw=none,name path=png_lb] table[col sep=comma,x index=0,y index=2] {data/latency_snr_png.dat};
        \addplot[draw=none,name path=png_ub] table[col sep=comma,x index=0,y index=3] {data/latency_snr_png.dat};
        \addplot[png_color,opacity=0.07] fill between[of=png_lb and png_ub];
        
        \addplot[jpeg_color,thick,densely dashed] table[col sep=comma,x index=0,y index=1] {data/latency_snr_jpeg.dat};
        \addplot[draw=none,name path=jpeg_lb] table[col sep=comma,x index=0,y index=2] {data/latency_snr_jpeg.dat};
        \addplot[draw=none,name path=jpeg_ub] table[col sep=comma,x index=0,y index=3] {data/latency_snr_jpeg.dat};
        \addplot[jpeg_color,opacity=0.07] fill between[of=jpeg_lb and jpeg_ub];

        \addplot[domain=0:10,rgb_color,thick,densely dotted] {48.56000};
        \addplot[domain=0:10,draw=none,name path=rgb_lb] {46.42000};
        \addplot[domain=0:10,draw=none,name path=rgb_ub] {50.77000};
        \addplot[rgb_color,opacity=0.07] fill between[of=rgb_lb and rgb_ub];

        \addplot[domain=0:10,vqvae_color,thick] {1.04000};
        \addplot[domain=0:10,draw=none,name path=vqvae512_lb] {0.75000};
        \addplot[domain=0:10,draw=none,name path=vqvae512_ub] {1.39000};
        \addplot[vqvae_color,opacity=0.07] fill between[of=vqvae512_lb and vqvae512_ub];

        \addplot[domain=0:10,vqvae_color,thick,densely dashdotdotted] {0.36000};
        \addplot[domain=0:10,draw=none,name path=vqvae8_lb] {0.20000};
        \addplot[domain=0:10,draw=none,name path=vqvae8_ub] {0.58000};
        \addplot[vqvae_color,opacity=0.07] fill between[of=vqvae8_lb and vqvae8_ub];
    \end{semilogyaxis}
    \end{tikzpicture}
    \label{fig:latency-SNR}
    }%
    \subfloat[$C=5$, $\gamma=6$~dB]{%
    \begin{tikzpicture}
    \begin{semilogyaxis}[
        width=0.33\linewidth,
        height=0.25\linewidth,
        xlabel={Channel non-access probability, $p$},
        ylabel={Latency, $L$ (s)},
        xmin={0}, xmax={1}, xtick distance={0.2},
        ymin={1e-2}, ymax={1e3}, try min ticks log={6},
    ]
        \addplot[png_color,thick,densely dashdotted] table[col sep=comma,x index=0,y index=1] {data/latency_p_png.dat};
        \addplot[draw=none,name path=png_lb] table[col sep=comma,x index=0,y index=2] {data/latency_p_png.dat};
        \addplot[draw=none,name path=png_ub] table[col sep=comma,x index=0,y index=3] {data/latency_p_png.dat};
        \addplot[png_color,opacity=0.07] fill between[of=png_lb and png_ub];
        
        \addplot[jpeg_color,thick,densely dashed] table[col sep=comma,x index=0,y index=1] {data/latency_p_jpeg.dat};
        \addplot[draw=none,name path=jpeg_lb] table[col sep=comma,x index=0,y index=2] {data/latency_p_jpeg.dat};
        \addplot[draw=none,name path=jpeg_ub] table[col sep=comma,x index=0,y index=3] {data/latency_p_jpeg.dat};
        \addplot[jpeg_color,opacity=0.07] fill between[of=jpeg_lb and jpeg_ub];

        \addplot[rgb_color,thick,densely dotted] table[col sep=comma,x index=0,y index=1] {data/latency_p_rgb.dat};
        \addplot[draw=none,name path=rgb_lb] table[col sep=comma,x index=0,y index=2] {data/latency_p_rgb.dat};
        \addplot[draw=none,name path=rgb_ub] table[col sep=comma,x index=0,y index=3] {data/latency_p_rgb.dat};
        \addplot[rgb_color,opacity=0.07] fill between[of=rgb_lb and rgb_ub];

        \addplot[vqvae_color,thick] table[col sep=comma,x index=0,y index=1] {data/latency_p_vqvae512.dat};
        \addplot[draw=none,name path=vqvae512_lb] table[col sep=comma,x index=0,y index=2] {data/latency_p_vqvae512.dat};
        \addplot[draw=none,name path=vqvae512_ub] table[col sep=comma,x index=0,y index=3] {data/latency_p_vqvae512.dat};
        \addplot[vqvae_color,opacity=0.07] fill between[of=vqvae512_lb and vqvae512_ub];

        \addplot[vqvae_color,thick,densely dashdotdotted] table[col sep=comma,x index=0,y index=1] {data/latency_p_vqvae8.dat};
        \addplot[draw=none,name path=vqvae8_lb] table[col sep=comma,x index=0,y index=2] {data/latency_p_vqvae8.dat};
        \addplot[draw=none,name path=vqvae8_ub] table[col sep=comma,x index=0,y index=3] {data/latency_p_vqvae8.dat};
        \addplot[vqvae_color,opacity=0.07] fill between[of=vqvae8_lb and vqvae8_ub];
    \end{semilogyaxis}
    \end{tikzpicture}
    \label{fig:latency-p}
    }%
    \subfloat[$p=0.9$, $\gamma=6$~dB]{%
    \begin{tikzpicture}
    \begin{semilogyaxis}[
        width=0.33\linewidth,
        height=0.25\linewidth,
        xlabel={Number of channels, $C$},
        ylabel={Latency, $L$ (s)},
        xmin={1}, xmax={10}, xtick distance={1},
        ymin={5e-2}, ymax={2e2}, try min ticks log={4},
    ]
        \addplot[png_color,thick,densely dashdotted] table[col sep=comma,x index=0,y index=1] {data/latency_C_png.dat};
        \addplot[draw=none,name path=png_lb] table[col sep=comma,x index=0,y index=2] {data/latency_C_png.dat};
        \addplot[draw=none,name path=png_ub] table[col sep=comma,x index=0,y index=3] {data/latency_C_png.dat};
        \addplot[png_color,opacity=0.07] fill between[of=png_lb and png_ub];
        
        \addplot[jpeg_color,thick,densely dashed] table[col sep=comma,x index=0,y index=1] {data/latency_C_jpeg.dat};
        \addplot[draw=none,name path=jpeg_lb] table[col sep=comma,x index=0,y index=2] {data/latency_C_jpeg.dat};
        \addplot[draw=none,name path=jpeg_ub] table[col sep=comma,x index=0,y index=3] {data/latency_C_jpeg.dat};
        \addplot[jpeg_color,opacity=0.07] fill between[of=jpeg_lb and jpeg_ub];

        \addplot[rgb_color,thick,densely dotted] table[col sep=comma,x index=0,y index=1] {data/latency_C_rgb.dat};
        \addplot[draw=none,name path=rgb_lb] table[col sep=comma,x index=0,y index=2] {data/latency_C_rgb.dat};
        \addplot[draw=none,name path=rgb_ub] table[col sep=comma,x index=0,y index=3] {data/latency_C_rgb.dat};
        \addplot[rgb_color,opacity=0.07] fill between[of=rgb_lb and rgb_ub];

        \addplot[vqvae_color,thick] table[col sep=comma,x index=0,y index=1] {data/latency_C_vqvae512.dat};
        \addplot[draw=none,name path=vqvae512_lb] table[col sep=comma,x index=0,y index=2] {data/latency_C_vqvae512.dat};
        \addplot[draw=none,name path=vqvae512_ub] table[col sep=comma,x index=0,y index=3] {data/latency_C_vqvae512.dat};
        \addplot[vqvae_color,opacity=0.07] fill between[of=vqvae512_lb and vqvae512_ub];

        \addplot[vqvae_color,thick,densely dashdotdotted] table[col sep=comma,x index=0,y index=1] {data/latency_C_vqvae8.dat};
        \addplot[draw=none,name path=vqvae8_lb] table[col sep=comma,x index=0,y index=2] {data/latency_C_vqvae8.dat};
        \addplot[draw=none,name path=vqvae8_ub] table[col sep=comma,x index=0,y index=3] {data/latency_C_vqvae8.dat};
        \addplot[vqvae_color,opacity=0.07] fill between[of=vqvae8_lb and vqvae8_ub];
    \end{semilogyaxis}
    \end{tikzpicture}
    \label{fig:latency-C}
    }%
    \caption{Latency as a function of (a)~\gls{snr}, (b)~channel non-access probability, and (c)~the number of channels. Curves denote the median, while shaded regions represent the 1st and 99th percentiles.}
    \label{fig:latency}
\end{figure*}

In this section, we present numerical results to evaluate the latency and task performance of the proposed framework and to compare it with benchmark transmission schemes.\footnote{The source code used to generate these results is publicly available at \url{https://github.com/joaohis/SemCR}}

In the simulations, we assume image transmissions from the 160~px version of the Imagenette\footnote{\url{https://github.com/fastai/imagenette}} dataset, with 9469 images in the training split and 3925 in the test split, all centrally cropped to $128\times128$ pixels. All models are implemented in PyTorch~2.10.0 and optimized with the Adam~\cite{Kingma2015} algorithm. For the \gls{vq-vae}, the training batch size, learning rate, and commitment loss weight are set to 128, $3\times10^{-4}$, and 0.25, respectively. Moreover, two codebook sizes are considered so that $K\in\{8,512\}$, while the latent dimension is kept constant at $\Delta=64$. The \gls{vq-vae} encodes each $128 \times 128$ pixel image with an 8-bit color depth into $32 \times 32$ latent vectors, achieving a compression ratio of $b_0/b=384/\log_2 K$.

Regarding the communication parameters, we assume binary phase shift keying modulation ($M=2$), a bandwidth of $B=10$~MHz, and frames of $T_0=10$~ms, of which $T=9$~ms are used for data transmission and the rest for spectrum sensing. Each communication batch contains 100 images, while the number of channels, \gls{snr}, and channel non-access probability are swept such that $C\in[1,10]$, $\gamma\in[-6,15]$~dB, and $p\in[0,0.99]$.

As benchmarks, we consider transmission schemes that employ \gls{ldpc} channel coding with $(n,k)=(100,90)$ combined with lossless or lossy compression, or without source coding. \Gls{png} is adopted for lossless compression, achieving a compression ratio of 1.7 in the considered dataset. For lossy compression, \gls{jpeg} is considered with the quality parameter set per image to achieve a compression ratio of 40. Lastly, for the case without source coding, we assume the transmission of raw RGB image data. Retransmissions are enabled only for the methods that employ \gls{png} or \gls{jpeg}. We start by discussing the latency results.

\subsection{Latency Evaluation}

Figs.~\ref{fig:latency-SNR}, \ref{fig:latency-p}, and~\ref{fig:latency-C} depict the communication latency as a function of the \gls{snr}, the channel non-access probability, and the number of channels, respectively. These results reveal that the proposed framework achieves substantial latency gains compared to all the considered benchmarks. The gains are more pronounced with a smaller codebook, as fewer bits are required to represent the indices of the latent vectors. The latency benefits of the proposed scheme remain consistent with different spectrum availability, as shown in Figs.~\ref{fig:latency-p} and \ref{fig:latency-C}.

In Fig.~\ref{fig:latency-SNR}, we observe that the latency of the benchmarks that employ \gls{png} and \gls{jpeg} increases rapidly as \gls{snr} drops below 7~dB, driven by a surge in retransmissions due to block errors. This does not occur in the proposed framework or the raw-RGB benchmark, where latency is independent of \gls{snr} because retransmissions are disabled. As shown later, this latency benefit comes at the cost of increased distortion and degraded task accuracy. Furthermore, for \gls{snr} above 7~dB, the latency of the \gls{jpeg} benchmark approaches that of \gls{vq-vae}(512), due to their similar compression ratios of 40 and 42.66, respectively. Despite this, none of the considered benchmarks outperforms the low latency of the proposed framework with a codebook size of 8, which achieves a compression ratio of 128. Especially, Fig.~\ref{fig:latency-SNR} reveals that \gls{vq-vae}(8) achieves latency gains of at least $134\times$, $79\times$, and $3.3\times$ compared to the raw-RGB, \gls{png}, and \gls{jpeg} benchmarks, respectively.

\subsection{Task Performance Evaluation}

\begin{figure}[t]
    \centering
    \begin{tikzpicture}
    \begin{axis}[
        width=\linewidth,
        height=0.5\linewidth,
        xlabel={SNR, $\gamma$ (dB)},
        ylabel={Task accuracy},
        xmin={-6}, xmax={15}, xtick distance={2},
        ymin={0.7}, ymax={0.95}, ytick distance={0.05},
        yticklabel style={/pgf/number format/fixed zerofill},
        legend pos=south east,
        legend style={fill=white,fill opacity=0.7,text opacity=1,legend cell align=left,column sep=5pt},
        legend image post style={opacity=1},
    ]
        \addplot[vqvae_color,thick,opacity=0.3,forget plot] table[col sep=comma,x index=0,y index=1] {data/accuracy_vqvae512.dat};
        \addplot[only marks, mark=*,vqvae_color,forget plot] table[col sep=comma,x index=0,y index=1] {data/accuracy_vqvae512.dat};
    
        \addplot[,vqvae_color,thick,densely dashdotdotted,opacity=0.3,forget plot] table[col sep=comma,x index=0,y index=1] {data/accuracy_vqvae8.dat};
        \addplot[only marks, mark=triangle*,vqvae_color,forget plot] table[col sep=comma,x index=0,y index=1] {data/accuracy_vqvae8.dat};

        \addplot[png_color,thick,densely dashdotted,opacity=0.3,forget plot] table[col sep=comma,x index=0,y index=1] {data/accuracy_png.dat};
        \addplot[only marks, mark=square*,png_color,forget plot] table[col sep=comma,x index=0,y index=1] {data/accuracy_png.dat};

        \addplot[jpeg_color,thick,densely dashed,opacity=0.3,forget plot] table[col sep=comma,x index=0,y index=1] {data/accuracy_jpeg.dat};
        \addplot[only marks, mark=diamond*,jpeg_color,forget plot] table[col sep=comma,x index=0,y index=1] {data/accuracy_jpeg.dat};

        \addplot[rgb_color,thick,densely dotted,opacity=0.3,forget plot] table[col sep=comma,x index=0,y index=1] {data/accuracy_rgb.dat};
        \addplot[only marks, mark=pentagon*,rgb_color,forget plot] table[col sep=comma,x index=0,y index=1] {data/accuracy_rgb.dat};

        \addlegendimage{only marks,vqvae_color,mark=*}
        \addlegendentry{Proposed: VQ-VAE(512)}
        \addlegendimage{only marks,vqvae_color,mark=triangle*}
        \addlegendentry{Proposed: VQ-VAE(8)}
        \addlegendimage{only marks,png_color,mark=square*}
        \addlegendentry{PNG+LDPC(100,90)}
        \addlegendimage{only marks,jpeg_color,mark=diamond*}
        \addlegendentry{JPEG+LDPC(100,90)}
        \addlegendimage{only marks,rgb_color,mark=pentagon*}
        \addlegendentry{Raw RGB+LDPC(100,90)}
    \end{axis}
    \end{tikzpicture}
    \caption{Task accuracy as a function of \gls{snr}. For the benchmarks that employ \gls{png} and \gls{jpeg}, the number of retransmissions is limited to 2.}
    \label{fig:accuracy}
\end{figure}
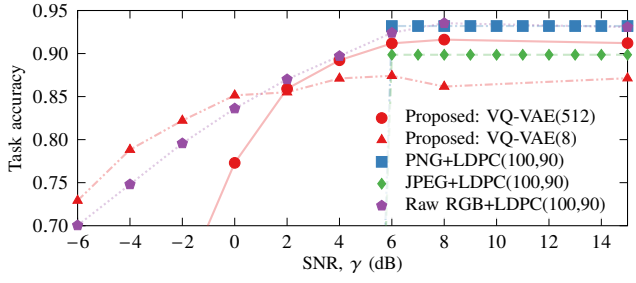

We now evaluate the task accuracy achieved by the \gls{crx} for the proposed framework and benchmarks. For the downstream task, we consider image classification by a MobileNetV3-Large model~\cite{Howard2019}. In the implementation, we adopt a transfer learning approach by initializing the model with weights pre-trained on a large-scale dataset and, for each scenario, fine-tuning the model on the received images of the target dataset. Fine-tuning is performed with a training batch size of $64$ and a learning rate of $10^{-3}$. The performance metric captures both the receiver's ability to decode images, even with distortion, and the resulting classification accuracy. Thus, the task accuracy is zero when an image cannot be decoded and equals the classification accuracy otherwise.

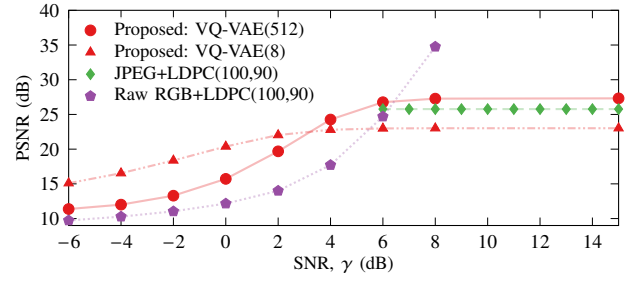
\begin{figure}[t]
    \centering
    \begin{tikzpicture}
    \begin{axis}[
        width=\linewidth,
        height=0.5\linewidth,
        xlabel={SNR, $\gamma$ (dB)},
        ylabel={PSNR (dB)},
        xmin={-6}, xmax={15}, xtick distance={2},
        ymin={9}, ymax={40}, ytick distance={5},
        legend pos=north west,
        legend style={fill=white,fill opacity=0.7,text opacity=1,legend cell align=left,column sep=5pt},
        legend image post style={opacity=1},
    ]
        \addplot[vqvae_color,thick,opacity=0.3,forget plot] table[col sep=comma,x index=0,y index=1] {data/distortion_vqvae512.dat};
        \addplot[only marks, mark=*,vqvae_color,forget plot] table[col sep=comma,x index=0,y index=1] {data/distortion_vqvae512.dat};
    
        \addplot[,vqvae_color,thick,densely dashdotdotted,opacity=0.3,forget plot] table[col sep=comma,x index=0,y index=1] {data/distortion_vqvae8.dat};
        \addplot[only marks, mark=triangle*,vqvae_color,forget plot] table[col sep=comma,x index=0,y index=1] {data/distortion_vqvae8.dat};

        \addplot[jpeg_color,thick,densely dashed,opacity=0.3,forget plot] table[col sep=comma,x index=0,y index=1] {data/distortion_jpeg.dat};
        \addplot[only marks, mark=diamond*,jpeg_color,forget plot] table[col sep=comma,x index=0,y index=1] {data/distortion_jpeg.dat};

        \addplot[rgb_color,thick,densely dotted,opacity=0.3,forget plot] table[col sep=comma,x index=0,y index=1] {data/distortion_rgb.dat};
        \addplot[only marks, mark=pentagon*,rgb_color,forget plot] table[col sep=comma,x index=0,y index=1] {data/distortion_rgb.dat};

        \addlegendimage{only marks,vqvae_color,mark=*}
        \addlegendentry{Proposed: VQ-VAE(512)}
        \addlegendimage{only marks,vqvae_color,mark=triangle*}
        \addlegendentry{Proposed: VQ-VAE(8)}
        \addlegendimage{only marks,jpeg_color,mark=diamond*}
        \addlegendentry{JPEG+LDPC(100,90)}
        \addlegendimage{only marks,rgb_color,mark=pentagon*}
        \addlegendentry{Raw RGB+LDPC(100,90)}
    \end{axis}
    \end{tikzpicture}
    \caption{\gls{psnr} of the successfully decoded images as a function of \gls{snr}.}
    \label{fig:distortion}
\end{figure}

Fig.~\ref{fig:accuracy} shows the task accuracy as a function of \gls{snr}, while Fig.~\ref{fig:distortion} depicts the \gls{psnr} of the successfully decoded images as a function of \gls{snr}. To limit latency in this evaluation, we set the maximum number of retransmissions per block to 2 for the \gls{png} and \gls{jpeg} benchmarks. Under this constraint, these schemes cannot decode the images when \gls{snr} is below 6~dB, resulting in a zero task accuracy.\footnote{In practice, link adaptation may restore communication by selecting a lower coding rate~($k/b$), at the cost of additional latency due to feedback exchange and the transmission of additional parity bits.} This does not occur in the proposed framework or in the raw-RGB benchmark, where task performance and distortion degrade gracefully as \gls{snr} decreases, allowing task execution even under channel fading. When \gls{snr} is above 6~dB, the proposed framework with a codebook size of 512 achieves task accuracy comparable to standard separate source and channel coding schemes, with a marginal drop of up to 2\% relative to the \gls{png} benchmark, and a gain of up to 1.7\% over that with \gls{jpeg}. Moreover, for \gls{snr} below 2~dB, the proposed framework achieves higher task accuracy with a smaller codebook, thanks to the lower \gls{bler} obtained with shorter blocks. Lastly, Fig.~\ref{fig:distortion} highlights that the proposed framework also exhibits reasonable image reconstruction performance, indicating its potential for other computer vision tasks.

Considering the latency and task accuracy results jointly, we note that the proposed framework achieves low-latency communication that scales well under limited spectrum availability. At the same time, it maintains competitive task accuracy, outperforming conventional source coding schemes by enabling task execution even under challenging channel conditions.

\section{Conclusion} \label{sec:conclusion}

In this paper, we proposed a framework for task-oriented, low-latency image transmission with opportunistic spectrum access.
By leveraging discrete, low-dimensional latent representations learned via a \gls{vq-vae}, the framework reduces the number of transmitted bits while remaining compatible with conventional digital communication pipelines. Numerical results demonstrate a favorable latency-accuracy trade-off, with substantial latency reductions and only marginal accuracy degradation compared to standard benchmarks. Overall, the proposed framework enables efficient operation under limited spectrum availability by reducing the number of transmitted bits and avoiding retransmissions, while supporting uninterrupted task execution even under adverse channel conditions.

%
%
\appendices

\section{Encoder and Decoder Model Architectures}

The implemented encoder and decoder architectures are listed from input to output in Table~\ref{tab:architecture}. In the table, ``Conv2d'' denotes a two-dimensional convolutional layer, whereas ``ConvTranspose2d'' denotes its transposed counterpart. ``ResidualStack'' denotes a stack of residual blocks following the ResNet architecture. All layers use the rectified linear unit as the activation function, denoted as ``ReLU''. Further details on the hyperparameters can be found in the PyTorch~2.10.0 documentation and the source code provided with this work.

\begin{table*}[t]
    \caption{Encoder and Decoder Model Architectures}
    \label{tab:architecture}
    \centering
    \begin{tabular}{cllcc}
        \toprule
        \textbf{Model} & \textbf{Layer type} & \textbf{Hyperparameters} & \textbf{Activation} & \textbf{Output channels}\\
        \midrule
        \multirow{8}{*}{Encoder} & Conv2d & \texttt{kernel\_size=4}, \texttt{stride=2}, \texttt{padding=1} & ReLU & 64\\
        & Conv2d & \texttt{kernel\_size=4}, \texttt{stride=2}, \texttt{padding=1} & ReLU & 128\\
        & Conv2d & \texttt{kernel\_size=3}, \texttt{stride=1}, \texttt{padding=1} & ReLU & 128\\
        \cmidrule{2-5}
        & \multirow{3}{*}{ResidualStack} & \texttt{num\_blocks=2}, \texttt{mid\_channels=32} & \multirow{3}{*}{ReLU} & \multirow{3}{*}{128}\\
        && \textit{Layer 1:} \texttt{kernel\_size=3}, \texttt{stride=1}, \texttt{padding=1} &&\\
        && \textit{Layer 2:} \texttt{kernel\_size=1}, \texttt{stride=1}, \texttt{padding=0} &&\\
        \cmidrule{2-5}
        & Conv2d & $\texttt{kernel\_size=1}$, \texttt{stride=1}, \texttt{padding=0} & ReLU & $\Delta$\\
        \midrule
        \multirow{7}{*}{Decoder} & ConvTranspose2d & \texttt{kernel\_size=4}, \texttt{stride=2}, \texttt{padding=1} & ReLU & 128\\
        \cmidrule{2-5}
        & \multirow{3}{*}{ResidualStack} & \texttt{num\_blocks=2}, \texttt{mid\_channels=32} & \multirow{3}{*}{ReLU} & \multirow{3}{*}{128}\\
        && \textit{Layer 1:} \texttt{kernel\_size=3}, \texttt{stride=1}, \texttt{padding=1} &&\\
        && \textit{Layer 2:} \texttt{kernel\_size=1}, \texttt{stride=1}, \texttt{padding=0} &&\\
        \cmidrule{2-5}
        & ConvTranspose2d & \texttt{kernel\_size=4}, \texttt{stride=2}, \texttt{padding=1} & ReLU & 64\\
        & ConvTranspose2d & \texttt{kernel\_size=4}, \texttt{stride=2}, \texttt{padding=1} & ReLU & 3\\
        \bottomrule
    \end{tabular}
\end{table*}

\begin{figure*}[t]
    \centering
    \includegraphics[width=\linewidth]{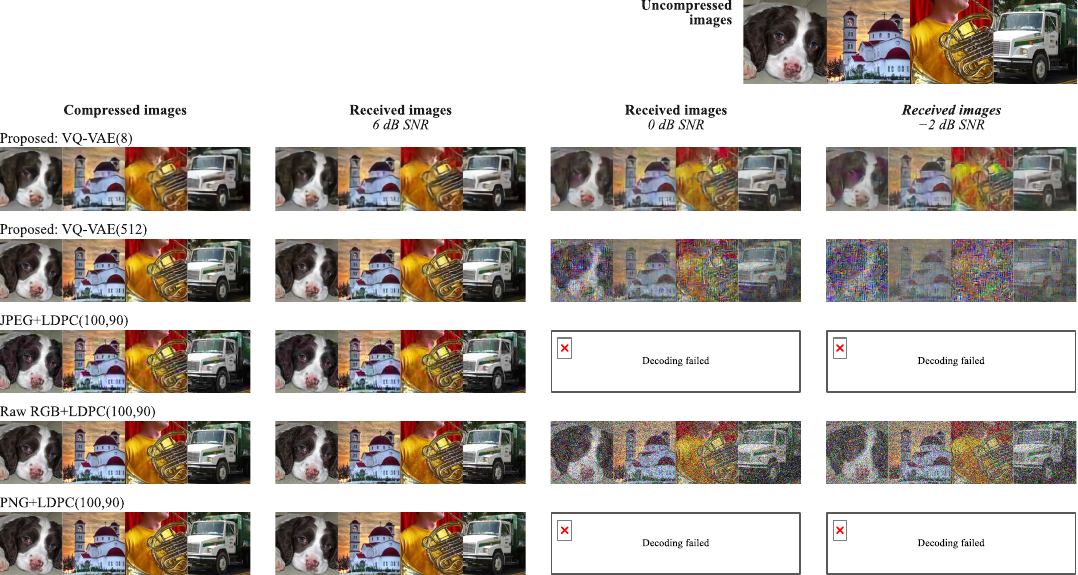}
    \caption{Examples of compressed images by the \gls{ctx} and received images by the \gls{crx} considering the proposed framework and the benchmark transmission schemes. For the benchmarks that employ \gls{png} and \gls{jpeg}, the number of retransmissions is limited to 2.}
    \label{fig:qualitative-results}
\end{figure*}

\section{Qualitative Evaluation}

To illustrate the impact of compression and communication errors on image quality, Fig.~\ref{fig:qualitative-results} presents examples of compressed and received images obtained with the proposed framework and the benchmark transmission schemes. The label ``Decoding failed'' indicates cases where the received \gls{jpeg} and \gls{png} streams could not be decoded due to bitstream corruption, assuming a maximum of 2 retransmissions per block.

The examples show that images compressed by \gls{vq-vae} exhibit smoother edges and a loss of fine detail, both of which become more pronounced as the codebook size decreases. Nevertheless, \gls{vq-vae} largely preserves the shapes, edges, and colors of the original image and avoids the block-boundary artifacts observed with \gls{jpeg} compression. Additionally, consistent with the trend in Fig.~\ref{fig:distortion}, received images using \gls{vq-vae} and raw RGB exhibit increasing distortion as the \gls{snr} decreases due to erroneously received blocks. For \gls{vq-vae} specifically, reception errors produce distinct degradation patterns across images. For instance, at an \gls{snr} of $-2$~dB, the brass-player image compressed with \gls{vq-vae}(512) is severely degraded by grain-like artifacts, whereas the truck image largely retains its original shape. At the same \gls{snr}, \gls{vq-vae}(8) demonstrates greater robustness, as the dog, church, and truck images preserve their shapes and edges, exhibiting mainly color distortions.

\balance

%
%
\bibliographystyle{IEEEtran}
\bibliography{IEEEabrv,references}

\end{document}